\documentclass[submit]{smj}

 \usepackage{appendix}
 \usepackage{booktabs}
 \usepackage{rotating}
\usepackage{algorithm}
\usepackage{algpseudocode}
\usepackage{array}
\newcolumntype{H}{>{\setbox0=\hbox\bgroup}c<{\egroup}@{}}
 
 \def \IR{\hbox{{\rm I}\kern-.2em\hbox{{\rm R}}}}
 \newcommand{\mv}[1]{{\boldsymbol{\mathrm{#1}}}}

 \DeclareMathOperator{\E}{\mathbb{E}}

 \graphicspath{{figs/}{./}}

\Author{\"{O}zg\"{u}r Asar\Affil{1}}
\AuthorRunning{\"{O}zg\"{u}r Asar}

\Affiliations{

\item Department of Biostatistics and Medical Informatics, 
      Faculty of Medicine, 
      Ac{\i}badem Mehmet Ali Ayd{\i}nlar University,
      \.{I}stanbul, Turkey

}   

\CorrAddress{\"{O}zg\"{u}r Asar, 
             Department of Biostatistics and Medical Informatics, 
             Faculty of Medicine,
             Ac{\i}badem Mehmet Ali Ayd{\i}nlar University, 
             Kerem Ayd{\i}nlar Kamp\"{u}s\"{u}, 
             Kay{\i}\c{s}da\u{g}{\i} Cad., No: 32, 
             Ata\c{s}ehir, \.{I}stanbul, Turkey}
\CorrEmail{ozgur.asar@acibadem.edu.tr $|$ ozgurasarstat@gmail.com}
\CorrPhone{(+90)\;216\;500\;4255}
\CorrFax{(+90)\;216\;576\;5120}

\Title{Bayesian analysis of Turkish 
 Income and Living Conditions data, using clustered 
 longitudinal ordinal modelling with Bridge distributed 
 random-effects}
\TitleRunning{Bayesian analysis of Turkish Income and Living Conditions data}

\Abstract{
 This paper is motivated by the panel surveys, 
 called Statistics on Income and Living Conditions (SILC), 
 conducted annually on 
 (randomly selected) country-representative households 
 to monitor EU 2020 aims on poverty reduction.  
 We particularly consider the surveys conducted in Turkey  
 within the scope of integration to the EU. 
 Our main interests are on health aspects of economic and living conditions.    
 The outcome is {\it self-reported health} that  
 is clustered longitudinal ordinal, 
 since repeated measures of it are nested within individuals and individuals are nested 
 within families. 
 Economic and living conditions were measured through a number of 
 individual- and family-level explanatory variables. 
 The questions of interest are 
 on the marginal relationships between the outcome 
 and covariates that we 
 address using a polytomous logistic regression with 
 Bridge distributed random-effects. 
 This choice of distribution allows 
 us to {\it directly} obtain marginal inferences 
 in the presence of random-effects. 
 Widely used Normal distribution is also considered 
 as the random-effects distribution. 
 Samples from the joint posterior densities of 
 parameters and random-effects are drawn
 using Markov Chain Monte Carlo. 
 Interesting findings from public health point of view are that  
 differences were found between sub-groups of 
 employment status, income level and panel year  
 in terms of odds of reporting better health.
}


\begin{document}

\maketitle

\section{Introduction}

 Statistics on Income and Living Conditions (SILC) 
 are panel surveys that have been conducted in 28 EU and 4 non-EU 
 countries (Turkey, Iceland, Switzerland and Norway)
 to monitor EU 2020 strategies on poverty reduction. 
 In SILC, households are followed annually.  
 Detailed information on 
 income, poverty, social exclusion, 
 living conditions, housing, labour, education and health 
 are collected by questionnaires.  
 The countries are not expected to use 
 the same questionnaires. They can do modifications on the 
 main questionnaire based on local conditions 
 as long as they collect the 
 minimum information.   

 In SILC, main study units are households. 
 Data from a single country has 
 a three-level data structure, 
 since repeated observations are nested 
 within individuals and individuals are nested 
 within families. This kind of structure can 
 also be called as clustered longitudinal. 
 Income and living conditions have been measured 
 through a number of individual- and family-level 
 variables, e.g. mean household disposable income, 
 gender, marital status, age, education level and 
 working status.  
 Health status is measured through 
 self-reported health (SRH): individuals' rate to the question, 
 ``How is your health in general?". The rates can be one 
 of the followings: very bad, bad, fair, good, and very good. 
 SRH is an important 
 indicator of individuals' general health and argued to be a 
 good predictor of morbidity and mortality \citep{burstrom2001}. 
 
 In the current study, we are interested in exploring 
 the relationships between health status and income 
 and living conditions. 
 We obtained data from the Turkish SILC (TR-SILC).  
 It has been conducted by 
 Turkish Statistical Institute (TURKSTAT) 
 within the scope of European Union Statistics on Income and 
 Living Conditions (EU-SILC).   
 There are a number of papers that analysed EU-SILC 
 data, excluding TR-SILC. 
 To the best of our knowledge, 
 there is no work that analysed panels from TR-SILC. 
 Detailed literature review on SILC is provided 
 in Section \ref{sec:literature}. 
 None of the works considered drawing marginal 
 inference (to be introduced below) 
 with appropriate statistical modelling. 

 Our scientific questions are on the interpretations 
 of the relationships between SRH and 
 economic and demographic variables. 
 Therefore, our first natural choice would be 
 working with marginal models \citep[Chapter~8]{diggle2002}. 
 For inferential purposes, generalised estimating equations 
 (GEE; \cite{liang1986}) could be used. 
 However, this method does not work with a 
 genuine likelihood function, and might not 
 be the best option for unbalanced data. 
 Instead, we consider random-effects models (Chapter 10 of 
 \cite{diggle2002}).  
 This class of model consists of individual-level 
 terms together with covariates. 
 Interpretations of the regression coefficients are 
 typically based on the assumption that 
 two persons belonging to different covariate sub-groups 
 have the same individual characteristics.  
 However, this would be unrealistic in many cases.  
 \cite{wang2003, wang2004} invented a class of distribution, 
 called {\it Bridge} distribution for logit link, 
 that allows obtaining marginal 
 interpretations for the covariates 
 within the random-effects modelling framework.   
 So far, the distribution is used only for binary data, 
 mostly within the scope of analysis of 
 longitudinal data, i.e. two-level 
 data \citep{bandyopadhyay2010, parzen2011, tu2011, li2011}.  
 \cite{tom2016} used it for semi-continuous 
 outcome data in the binary logistic sub-part model. 
 \cite{boehm2013} considered Bridge distribution for 
 multi-level spatial binary data. 
 In this study, we use Bridge distribution for analysis 
 of three-level ordinal outcome data. 
 We take the Bayesian paradigm for inference. 
 The No-U-Turn Sampler \citep{hoffman2014}, 
 an adaptive version of Hamiltonian Monte Carlo \citep{neal2011},  
 is used to obtain samples from the joint posterior distributions  
 of parameters and random-effects. 
 
 Rest of the paper is organised as follows. 
 In Section \ref{sec:data}, we give the details 
 of TR-SILC data. 
 In Section \ref{sec:literature}, we review 
 the literature on both longitudinal ordinal 
 modelling and analysis of SILC data.    
 Section \ref{sec:model} presents model 
 formulation and inferential details. 
 Section \ref{sec:silc_analysis} presents 
 results on the TR-SILC data-set. 
 We close the paper by conclusion and discussion. 

 \section{TR-SILC Data-set}
 \label{sec:data}
 
 \subsection{Background information} 
 
 SILC surveys 
 have been conducted in 28 EU countries and 
 4 non-EU countries (Turkey, Iceland, Switzerland and Norway)  
 to monitor EU 2020 strategies on poverty reduction.  
 The surveys cover objective and subjective questions on both 
 monetary and non-monetary aspects  
 of income, social inclusion and living conditions. 
 Family- and individual-level micro-data 
 are collected on income, poverty, social inclusion, 
 living conditions, housing, labour, 
 education and health. The surveys have been conducted as 
 cross-sectionally and also as panels of four years. 
 For more details, interested reader is referred to 
 a data-resource paper on EU-SILC by \cite{arora2015} and to 
 the website of Eurostat at
 \begin{center}
 \url{http://ec.europa.eu/eurostat/web/income-and-living-conditions}.
 \end{center}
 The surveys have been conducted in Turkey by TURKSTAT 
 starting from 2006 within the scope of integration to the EU. 
 Since then data have been collected annually as cross-sections 
 and panels of four years. Country-representative families 
 have been randomly selected, and those willing to participate 
 have been included. Every year of the panels, 
 new families (and individuals from these families), 
 and/or new individuals, 
 e.g. newborns, are included. 
 If an individual leaves an existing family, and forms a new one, 
 e.g. by getting married,   
 the new family is also included in the study. 
 For such a person, 
 identity number is kept the same, and 
 the new family is assigned a new identity number. 
 More details regarding TR-SILC could be found at
 \begin{center} 
 \url{http://www.turkstat.gov.tr/PreTablo.do?alt_id=1011}.
 \end{center} 
 
 \subsection{Data-set}

 In the current study, we consider the panel of 2010 -- 2013.
 Subjects who are older than 16 (inclusive) are considered, since 
 SRH is not available for those who are younger than 16. 
 There are a total of 109 066 records in the available data-set. 
 Summary statistics on number of records, 
 and family-level variables are displayed 
 in Table \ref{tab:summary_stats1}, whereas summary statistics on 
 individual-level variables are displayed in Table \ref{tab:summary_stats2}. 
 Every year, new families 
 and individuals were 
 recruited to the panel. 
 For example, whilst there were 3 056 
 families (8 090 individuals) available 
 at 2010, there were 8 712 
 families (23 391 individuals) at 2011. 
 The changes in the proportions for sub-groups of 
 family size, gender, marital status, age, 
 education level and working status were 
 less than a per cent across successive years.  
 Mean household disposable income (MHDI) is 
 calculated by dividing 
 household disposable income by family size. 
 Note that although we report categorised family 
 size in Table \ref{tab:summary_stats1}, 
 original family sizes were used for MHDI calculation. 
 MHDI was increasing through years, 
 e.g. the medians were 6 614, 7 061, 7 757 and 8 586 
 Turkish Lira for 2010, 2011, 2012 and 2013, respectively. 
 Clearly the variable is right-skewed, hence 
 log-transformation will be applied when it is 
 considered as an explanatory variable in a 
 statistical model. 
 
 Follow-up patterns for 
 families and individuals are 
 displayed in Table \ref{tab:patterns_silc}.   
 As can be seen, majority of the families and individuals 
 were present at 2012--2013, 2011--2012--2013 
 and 2010--2011--2012--2013. 
 There were 465 families and 
 corresponding 2 515 individuals 
 who were only present at 2013. 
 The rest corresponds to either drop-out 
 or intermittent missingness patterns, with 
 259 families and 873 individuals being in the latter pattern. 
 If an individual is available at a follow-up, 
 there is no missing data for her/him regarding 
 the variables reported 
 in Tables \ref{tab:summary_stats1} \& \ref{tab:summary_stats2}. 
 One might argue that missing-at-random 
 assumption \citep{little2002} would be reasonable 
 for the intermittent and drop-out patterns, 
 since reasons for missing data include moving to another country, 
 moving to nursing home, serving for the army, and so on. 
 Likelihood-based inference would be reliable 
 under this assumption \citep{diggle2002}. 
  
\begin{figure}[t]
	\centering
		\includegraphics[scale=0.75]{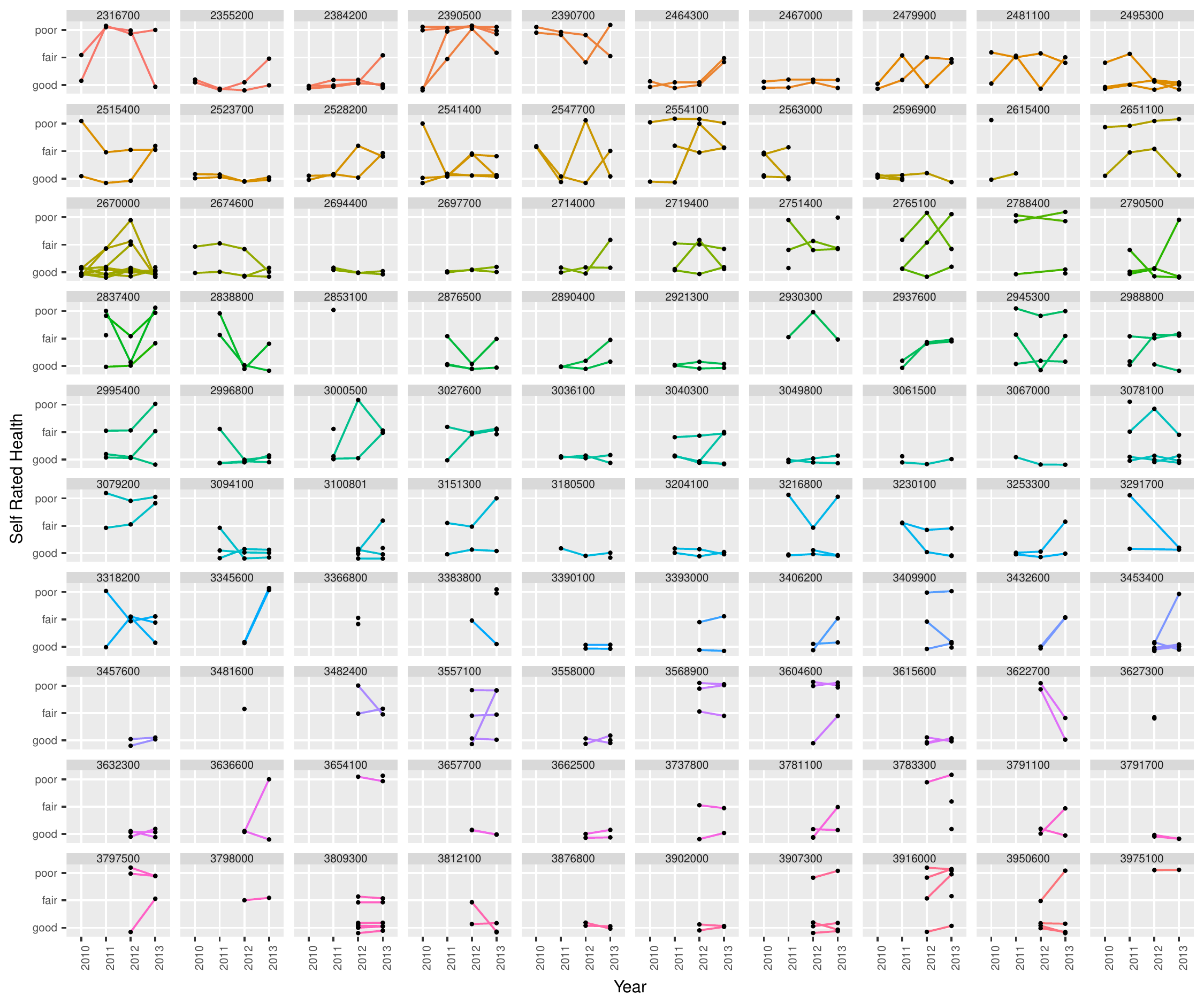}
	\caption{Spagetti-plots of SRH data for a random sample of 100 families and associated 284 individuals. 
	 SRH data are jittered for clearer representation.}
	\label{fig:spagetti_silc}
\end{figure}

 We consider re-categorising SRH as good 
 health (composed of good and very good), 
 fair health and poor 
 health (composed of bad and very bad) \citep{abebe2016, yardim2018}. 
 The percentage of people who reported good (bad) health 
 were increasing (decreasing) through 2010--2013. 
 There seems no clear pattern for fair health.  
  Spagetti-plots of SRH data for a random sample of 100 families 
 are displayed in Figure \ref{fig:spagetti_silc}. 
 In total, 699 repeated measures on 284 individuals from these 
 families are displayed. The plots indicate 
 that there are heterogeneities 
 both between families and between individuals 
 in terms of health status evolutions.

 \begin{table}[htbp]
 \begin{center}
 \caption{Summary statistics on number of records and family-level variables for the 2010 -- 2013 panel of TR-SILC. }
 \label{tab:summary_stats1}
 \fbox{
 \scalebox{0.8}{
 \begin{tabular}{l r r r r} 
                                         \multicolumn{5}{c}{{\bf Number of records}} \\ \hline
																         & 2010  & 2011   & 2012   & 2013 \\  \cmidrule(lr){2-5}
 {\it Families}      						    		 & 3 056 & 8 712  & 14 387 & 14 337 \\ 
 {\it Individuals}    						 			 & 8 090 & 23 391 & 38 733 & 38 852 \\ \hline
                                         \multicolumn{5}{c}{{\bf Family-level variables}} \\ \hline
																				 & 2010  & 2011   & 2012   & 2013 \\ \cmidrule(lr){2-5}
 {\it Family size}		                   & & & & \\
 1--2 	                                 & 1 766 (57.79\%) & 5 013 (57.54\%)  & 8 214 (57.09\%)  & 8 091 (56.43\%)\\
 3--5                                   & 1 196 (39.14\%) & 3 398 (39.00\%)  & 5 663 (39.36\%)  & 5 711 (39.83\%)\\
 6+                                      & 94 (3.08\%)     & 301   (3.46\%)   & 510  (3.54\%)    & 535   (3.73\%)\\[1ex] 
 \multicolumn{2}{l}{{\it Mean household disposable income (in Turkish Lira)}} & & & \\ 
 Min                                     & 545     & 0       & 62 & 6 \\
 25th           												 & 4 281   & 4 616   & 5 040 & 5 709 \\
 50th        													   & 6 614   & 7 061   & 7 757 & 8 586 \\
 Mean         													 & 9 360   & 9 678   & 10 452 & 11 556 \\
 75th         													 & 10 465  & 11 187  & 12 205 & 13 319 \\
 Max        													   & 427 450 & 232 720 & 285 686 & 373 925 \\ 
 Sd          													   & 13 212  & 10 450  & 10 922 & 12 001 \\
 \end{tabular}}}
 \end{center}
 \end{table}

\begin{table}[htbp]
 \begin{center}
 \caption{Summary statistics of individual-level variables 
 for the 2010--2013 panel of TR-SILC. }
 \label{tab:summary_stats2}
 \fbox{
 \scalebox{0.9}{
 \begin{tabular}{l r r r r} 
																			   & 2010  & 2011   & 2012   & 2013 \\ \cmidrule(lr){2-5}
 {\it Gender}			        		      		 & & & & \\ 
 Male           						  					 & 3 902 (48.23\%) & 11 277 (48.21\%) & 18 696 (48.27\%) & 18 741 (48.24\%) \\ 
 Female         						  					 & 4 188 (51.77\%) & 12 114 (51.79\%) & 20 037 (51.73\%) & 20 111 (51.76\%) \\[1ex] 
 {\it Marital status}	         	    		 & & & & \\
 Married           						  				 & 5 523 (68.27\%) & 15 960 (68.23\%) & 26 126 (67.45\%) & 26 080 (67.13\%) \\ 
 Never married         						  		 & 1 865 (23.05\%) & 5 447 (23.29\%)  & 9 156 (23.64\%)  & 9 298 (23.93\%)\\
 Widowed/separated             			     & 702 (8.68\%)    & 1 984 (8.48\%)   & 3 451 (8.91\%)   & 3 474 (8.94\%)\\[1ex] 
 {\it Age}			      					    		 & & & & \\ 
 16--34           											 & 3 215 (39.74\%) & 9 508 (40.65\%)  & 15 571 (40.20\%) & 15 476 (39.83\%) \\
 35--64        								 				 & 3 847 (47.55\%) & 11 059 (47.28\%) & 18 443 (47.62\%) & 18 613 (47.91\%) \\
 65+         													   & 1 028 (12.71\%) & 2 824 (12.07\%)  & 4 719 (12.18\%)  & 4 763 (12.26\%) \\[1ex] 
 {\it Education level}			       			 & & & & \\
 Primary/less                 				   & 4 614 (57.03\%) & 13 309 (56.90\%) & 21 571 (55.69\%) & 21 012 (54.08\%) \\
 Secondary/high            					     & 2 724 (33.67\%) & 7 886 (33.71\%)  & 13 370 (34.52\%) & 13 871 (35.70\%)\\  
 Higher education             					 & 752 (9.30\%)    & 2 196 (9.39\%)   & 3 792 (9.79\%)   & 3 969 (10.22\%)\\[1ex] 
 {\it Working status}		          			 & & & & \\
 Employed                   						 & 3 646 (45.07\%) & 10 622 (45.41\%) & 17 839 (46.06\%) & 17 802 (45.82\%) \\
 Unemployed        					             & 345   (4.26\%)  & 920 (3.93\%)     & 1 370 (3.54\%)   & 1 465 (3.77\%) \\ 
 Student                      					 & 656 (8.11\%)    & 1 809 (7.73\%)   & 3 027 (7.82\%)   & 3 220 (8.29\%) \\ 
 Retired                      					 & 678 (8.38\%)    & 1 909 (8.16\%)   & 3 096 (7.99\%)   & 3 109 (8.00\%) \\ 
 Housekeeper                      	   	 & 2 141 (26.46\%) & 6 353 (27.16\%)  & 10 309 (26.62\%) & 10 347 (26.63\%) \\ 
 Other                         					 & 624 (7.71\%)    & 1 778 (7.61\%)   & 3 092 (7.98\%)   & 2 909 (7.49\%) \\ [1ex]
 {\it Self reported health}  						 & & & & \\  
 Good                      					     & 5 151 (63.67\%) & 15 012 (64.18\%) & 25 516 (65.88\%) & 25 508 (65.65\%) \\
 Fair                       					   & 1 618 (20.00\%) & 4 931 (21.08\%)  &  7 734 (19.97\%) &  8 233 (21.19\%) \\ 
 Poor                          				   & 1 327 (16.33\%) & 3 448 (14.74\%)  & 5 483 (14.16\%)  &  5 111 (13.16\%)\\
 \end{tabular}}}
 \end{center}
 \end{table}

 \begin{table}[htbp]
 \begin{center}
 \caption{Follow-up patterns for families (F) and individuals (I) for the 2010 -- 2013 panel of TR-SILC. 
 $\checkmark$ indicates presence, $\times$ absence.}
 \label{tab:patterns_silc}
 \fbox{
 \scalebox{1}{
 \begin{tabular}{c c c c H r r} 
 2010        & 2011       & 2012       & 2013       & EF    & F & I \\ \hline
 \checkmark  & $\times$   & $\times$   & $\times$   & 71    & 71     & 333 \\
 $\times$    & \checkmark & $\times$   & $\times$   & 146   & 159    & 826 \\
 $\times$    & $\times$   & \checkmark & $\times$   & 259   & 329    & 1 389 \\
 $\times$    & $\times$   & $\times$   & \checkmark & 0     & 465    & 2 515 \\
 \checkmark  & \checkmark & $\times$   & $\times$   & 65    & 69     & 344 \\
 \checkmark  & $\times$   & \checkmark & $\times$   & 9     & 10     & 52 \\
 \checkmark  & $\times$   & $\times$   & \checkmark & 0     & 0      & 19 \\
 $\times$    & \checkmark & \checkmark & $\times$   & 234   & 259    & 1 113 \\
 $\times$    & \checkmark & $\times$   & \checkmark & 117   & 122    & 422 \\
 $\times$    & $\times$   & \checkmark & \checkmark & 5 383 & 5 655  & 15 388 \\
 \checkmark  & \checkmark & \checkmark & $\times$   & 86    & 87     & 430 \\
 \checkmark  & \checkmark & $\times$   & \checkmark & 47    & 48     & 147 \\
 \checkmark  & $\times$   & \checkmark & \checkmark & 80    & 79     & 252 \\
 $\times$    & \checkmark & \checkmark & \checkmark & 5 197 & 5 276  & 13 596 \\
 \checkmark  & \checkmark & \checkmark & \checkmark & 2 698 & 2 692  & 6 513 \\
 \end{tabular}}}
 \end{center}
 \end{table}

 \section{Literature Review}
 \label{sec:literature}
  
 \subsection{Modelling approaches for longitudinal ordinal data} 
 
 In this sub-section, we review papers that propose 
 modelling approaches for longitudinal ordinal data. 
 We selected papers that cover a wide range of approaches 
 to the topic. We first review papers on 
 longitudinal (two-level), then clustered longitudinal 
 (three-level) outcome.  
   
  \cite{kosorok1996} considered modelling ordinal data 
 in continuous time by letting the intensity matrix 
 being modified by explanatory variables. 
 Maximum likelihood (ML) was considered 
 for inference.  
 \cite{cowles1996} considered a tobit regression with 
 random-effects and used Gibbs sampling to sample 
 from the posterior densities.  
  \cite{varin2010} considered a polytomous probit  
 regression with a time-varying random-intercept that has an 
 exponential correlation structure. Parameters were 
 estimated by a pseudo-likelihood approach. 
  \cite{kaciroti2006} considered a polytomous logistic 
 regression with both random-effects and transition terms 
 within the context of 
 non-ignorable missing data. They used Bayesian 
 methods for inference. 
 \cite{burgin2015} proposed a mixed-effects 
 polytomous regression with tree-based 
 varying coefficients to take into account moderation 
 of the covariate effects. 
 ML was used for inference and 
 the {\tt R} package {\tt vcrpart} \citep{burgin2017} implements the 
 proposed methodology.        
 \cite{pulkstenis2001} considered a mixed-effects model 
 with log-log link and log-Gamma distributed random variables 
 that is coupled with a discrete survival model. 
 ML was used for inference.     
 \cite{laffont2014} considered a probit random-effects model 
 for multivariate longitudinal ordinal data. Stochastic 
 expectation-maximisation was considered for inference.  
 \cite{cheon2014} considered pattern-mixture type models with transition terms 
 and estimated the parameters using ML. 
 \cite{kauermann2000} considered a marginal polytoumous 
 logistic regression with time-varying regression 
 thresholds and covariate effects. Estimation was 
 carried out by a method called local estimation. 
 \cite{gadda2010} proposed a polytomous regression with 
 time-independent and time-dependent random-effects 
 with a smooth threshold term. Penalised likelihood 
 was used for inference.   

  \cite{lesaffre1996}, 
  \cite{molenberghs1997} and \cite{vansteen2001} 
  considered marginal modelling of longitudinal 
  ordinal data with the multivariate Dale model that is 
  coupled with a logistic regression to take into account 
  non-ignorable drop-out. The authors used ML was for inference.  
   \cite{heagerty1996} developed a modelling framework 
 that consists of two sub-models, a marginal polytomous regression 
 for the covariate effects and a global odds-ratio model for 
 pairwise association. GEE was considered for inference.   
 \cite{chen2014} extended their method to mis-classified response and covariate data. 
  \cite{ekholm2003} considered a polytomous marginal logistic regression 
 with dependence ratios and used ML  
 for inference.  
  \cite{perin2014} considered a marginal polytomous regression 
  and used GEE with orthogonalised 
 residuals for parameter estimation. 
   \cite{zaloumis2015} considered a marginal polytomous 
  regression with a non-proportional odds assumption 
  based on a multivariate logistic distribution 
  constructed by multivariate $t$ copula. 
  Bayesian methods were used for inference. 
  \cite{nooraee2016} considered a similar approach 
  by approximating 
  logistic distribution by  
  $t$ distribution with 8 degrees-of-freedom and used  
  ML for inference.   
   
 \cite{lee2007} and \cite{lee2008} considered 
 marginalised transition and random-effects 
 models for analysis of longitudinal ordinal data, 
 respectively. 
 \cite{lee2013} extended the marginalised random-effects 
 models to longitudinal bivariate ordinal outcomes. 
 \cite{lee2016} combined the transition and random-effects 
 terms in the marginalised modelling framework in order 
 to take into account long series of repeated measures. 
 All of these four works considered ML for inference.  
 \cite{rana2018} extended the work of \cite{lee2008} to 
 missing data and mis-reporting scenarios. 
  
   \cite{raman2005} and \cite{chan2015} considered polytomous logistic regression modelling 
 with random-effects for three-level ordinal outcomes.  
 \cite{liu2006} considered a three-level polytomous 
 logistic regression for longitudinal multivariate outcomes 
 within the item-response theory framework. 
 These works used ML for inference. 
  
 \subsection{Analysis of SILC data} 
 In this sub-section, we review the literature on analysis of SILC data 
 mainly from the statistical analysis point of view.  
 We found no work that considered analysis of SRH or 
 any panel aspect of the TR-SILC data. The works that 
 considered TR-SILC data are the followings. 
 \cite{oguzalper2015} derived variance estimators 
 for change in poverty rates using 2007 and 2008 cross-sections. 
 \cite{erus2015} investigated take-up of means-tested 
 health benefits using 2007 cross-section. 
 \cite{yardim2018} considered unmet access to 
 healthcare based on 2006 and 2013 
 cross-sections using separate multinomial 
 logistic regressions. 
 Therefore, we review in detail only the works that considered 
 analysis of the EU-SILC data. 
 In the papers reviewed, generally statistical 
 terminology is unclear and confusing. 
 We decided to exclude the papers with limited 
 information on statistical methods.  
 None of the works considered marginal inference 
 in the presence of random-effects. 
 Marginal inferences are based on fixed-effects 
 models that ignore dependency 
 due to the nested structure of the data which 
 is known to produce incorrect 
 uncertainty quantification.  

 \cite{vanderwel2011} considered impacts of sickness
 dimension of health on employment status using a cross-section   
 of data (year 2005) from 25 EU countries plus Norway and Iceland. 
 The authors used a mixed model for two-level binary 
 outcome (employed vs. unemployed) with a 
 country-level random intercept. 
 \cite{reeves2014} also considered the impacts of sickness
 dimension of health on employment status, 
 in the presence of the  
 2008 recession. 2006--2010 panel was divided into 
 two: 2006--2008 (pre-recession) and 2008--2010 (during recession). 
 Individual-level random 
 intercepts (ignoring family- and country-level characteristics) 
 were included in a mixed model for binary outcome. 
 \cite{ferraini2014} inspected the impacts of unemployment 
 insurance on dichotomised SRH in 23 EU countries 
 between 2006--2009, using a binary mixed model with 
 individual-level random intercept. 
 \cite{barlow2015} investigated through logistic regression the impacts 
 of austerity measures on SRH after the Recession 
 using the 2008--2011 Greek SILC data. 
 \cite{vaalavuo2016} considered the impacts 
 of unemployment and poverty on deterioration in SRH 
 from 26 EU countries using logistic regression. 
 \cite{pirani2015} inspected differences in health  
 between temporary and permanent contract working classes 
 using the 2007--2010 Italian SILC data. 
 The outcome is dichotomised SRH and 
 a marginal structural model was used. 
 \cite{huijts2015} inspected the impacts 
 of job loss during the Recession 
 using the 2009 EU-SILC data. 
 The authors used a linear mixed model for dichotomised health 
 with a country-level random intercept.  
 \cite{heggebo2015} inspected the effects of 
 health conditions on job loss during the 
 Recession using the 2007--2010 Scandinavian SILC data.  
 A linear mixed model with a individual-level 
 random intercept was used.  
 \cite{toge2015} analysed the 2008--2011 EU-SILC data 
 to understand the impacts of job loss on health. 
 They used a linear model for five-level SRH. 
 \cite{hessel2016} considered the impacts of 
 retirement on health using the 2009--2012 EU-SILC data. 
 The model was a linear mixed model with a individual-level random intercept. 
 \cite{toge2016} considered the impacts of 
 unemployment on SRH using the 2008--2011 EU-SILC data. 
 A linear mixed model with a subject-specific 
 random intercept was used. 
 \cite{abebe2016} inspected the impacts of 
 the Recession using the 2005--2011 EU-SILC data.  
 The authors used a mixed effects logistic 
 model for three-level SRH. 
 \cite{giannoni2016} considered 
 migrant health policies and health inequalities 
 using the 2012 EU-SILC data. 
 Ordinal SRH outcome was 
 analysed using a country-level random-intercept 
 in a polytomous regression. 
 \cite{clair2016} investigated the impacts 
 of housing payment problems on health 
 using the 2008--2010 EU-SILC data.  
 Dichotomised SRH was analysed using a 
 linear mixed effects model with 
 subject-, household- and country-level 
 random-effects. This is the only work that considered 
 family-level heterogeneity.  
 \cite{bacci2017} considered the 
 2009--2012 EU-SILC for investigating 
 impacts of economic 
 deprivation on SRH, using a 
 subject-specific random intercept 
 in a polytomous logistic regression 
 for ordinal SRH. 

 \section{Modelling framework}
 
 \label{sec:model}
 
 \subsection{Formulation}
 \label{sec:model_formulation} 

 Let $Y_{ijk}$ denote an ordinal response with $A$ possible values, i.e. $Y_{ijk} = 1, 2, \ldots, A$, 
 for subject $j \ (j = 1, \ldots, m_i )$ 
 belonging to family $i \ (i = 1, \ldots, n)$ at follow-up $k \ (k = 1, \ldots, s_{ij})$  
 at time $t_{ijk}$. Also let $\mv X_{ijk}$ denote the covariate matrix ($1 \times p$ dimensional) 
 attached to each of $Y_{ijk}$. There are $d_i = \sum_{j = 1}^{m_i} s_{ij}$ repeated measures for family $i$, 
 and $N = \sum_{i = 1}^{n} d_i$ follow-ups in total. 

 For TR-SILC, the target of inference is on the relationships between covariates and ordinal reponses, 
 i.e. on the conditional distribution of $[Y_{ijk} | \mv X_{ijk}, \mv \theta^m]$, where ``$[\cdot]$" stands for ``the distribution of", and 
 ${\mv \theta}^m$ are marginal parameters 
 in the sense that the relationship between covariates and 
 responses are not conditioned on other terms, e.g. response history and/or 
 individual characteristics.    
 A regression modelling framework for this distribution would be
 \begin{align}
 \label{eq:marginal_general}
 h \left\{ \mbox{P}(Y_{ijk} = a | {\mv X}_{ijk}, {\mv \theta}^m) \right\} = f(\mv X_{ijk}, {\mv \theta}^m), \ \ a = 1, \ldots, A - 1,
 \end{align}
 where $h(\cdot)$ is a link function, and 
 $f(\cdot)$ is a function that relates the covariates and associated 
 coefficients to the probability of ordinal outcome taking the 
 value $a$, $\mbox{P}(Y_{ijk} = a | {\mv X}_{ijk}, {\mv \theta}^m)$.   
 Choices of {\it cumulative logit} for $h(\cdot)$ and linear regression for $f(\cdot)$ would yield the 
 following polytomous logistic regression:
 \begin{align}
 \label{eq:marginal_cumlogit}
 logit \left\{ \mbox{P}(Y_{ijk} \leq a | {\mv X}_{ijk}, \alpha^m_a, {\mv \beta}^m_a) \right\} = 
 \alpha^m_a - {\mv X}_{ijk} {\mv \beta}^m_a.
 \end{align}
 Here, $\alpha^m_a$ is the category-specific intercept, 
 also known as the threshold, 
 ${\mv \beta}^m_a$ are the regression coefficients. 
 ${\mv \beta}^m_a$ address the target of inference for TR-SILC. 
 The issue with this model is that full-likelihood based inference 
 is not easy, since specification of the joint distribution of 
 ${\mv Y}_i = ({\mv Y}_{i1}^\top, \ldots, {\mv Y}_{im_i}^\top)^\top$, where ${\mv Y}_{ij} = (Y_{ij1}, \ldots, Y_{ijs_{ij}})^\top$,   
 a multivariate multinomial distribution with complex dependence structures, 
 is not straightforward in practice.  
 In order for likelihood based inference, one can include random-effects in \eqref{eq:marginal_cumlogit} 
 such that
 \begin{align}
 \label{eq:glmm_cumlogit}
 logit \left\{\mbox{P}(Y_{ijk} \leq a | {\mv X}_{ijk}, \alpha^c_a, {\mv \beta}^c_a, B_{ijk}) \right\} = \alpha^c_a - {\mv X}_{ijk} {\mv \beta}^c_a - B_{ijk},
 \end{align}
 where $B_{ijk}$ denotes the random-effect, 
 also known as latent-effect, for individual $j$ belonging to 
 family $i$ at the $k^{th}$ follow-up at time $t_{ijk}$. Note that $B_{ijk}$ is unobserved, 
 and typically a distribution is postulated to it. 
 The superscript $c$ in \eqref{eq:glmm_cumlogit} stands for the related terms being  
 conditional on $B_{ijk}$. The regression coefficients that address the target of inference, 
 $\mv \beta^m$, typically cannot be directly obtained from \eqref{eq:glmm_cumlogit} due to logit 
 link.  

 \subsection{Random-effects specifications}
 \label{sec:ref_distr}

A special, albeit useful approach would be to de-compose 
 the random effect term as $B_{ijk} = U_{i} + V_{ij}$, 
 where $U_{i} \bot V_{ij}$ \citep{raman2005, chan2015}. 
 This approach is also useful to take into account individuals in TR-SILC 
 who forms a new family but still included in the survey. 
 For such a patient, $U_{i}$ term would change, i.e. family characteristic would change, 
 but $V_{ij}$ would stay the same.   
 Assuming time-independent 
 random intercept terms, i.e. absence of the $k$ index in $U_{i}$ and $V_{ij}$ would 
 be sufficient to capture dependence for data-sets with 
 a few repeats per study-units. Note that for TR-SILC 
 majority of the families include at most five  
 individuals (see Table \ref{tab:summary_stats1}) and maximum 
 number of repeats per family/subject is four, 
 and there are many individuals 
 with less than four repeats (see Table \ref{tab:patterns_silc}).  
 
\begin{figure}[t]
	\centering
	\includegraphics[scale=0.5]{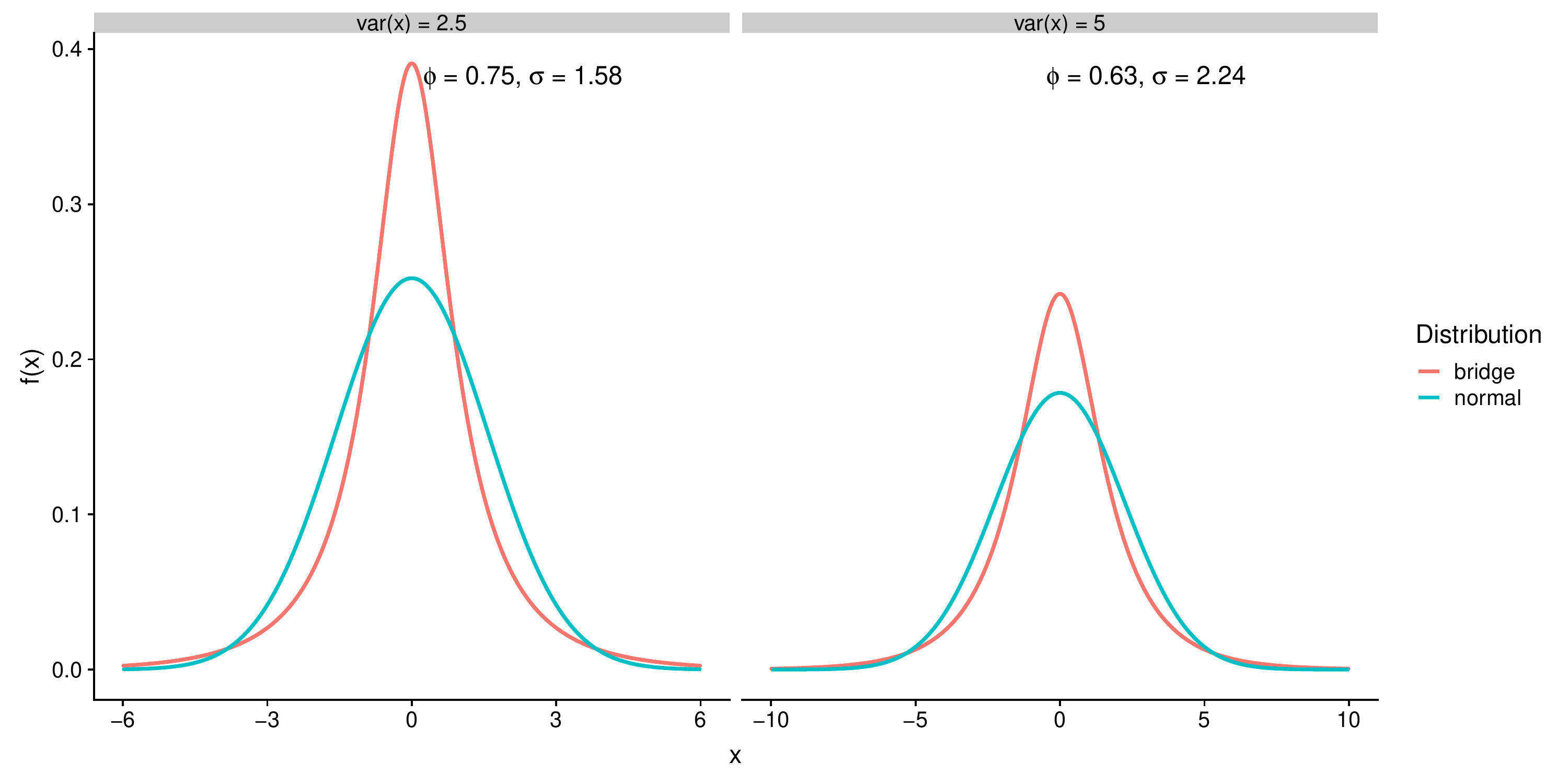}
	\caption{Density plots of Bridge and Normal distribution 
	for two settings of variance. 
	$\phi$ is the scale parameter of Bridge distribution, 
	$\sigma$ standard deviation of Normal.}
	\label{fig:distr}
\end{figure}

 The relationships between ${\mv \alpha}^m_a$ and ${\mv \alpha}^c_a$, 
 and ${\mv \beta}^m_a$ and ${\mv \beta}^c_a$, would be available 
 through solving the following convolution equation:
 \begin{equation}
 \mbox{P}(Y_{ijk} \leq a | \mv X_{ijk}, \alpha^m_a, \mv \beta^m_a) = 
\E_{U, V} \left\{ \mbox{P}(Y_{ijk} \leq a | \mv X_{ijk}, U_{i}, V_{ij}, \alpha^c_a, \mv \beta^c_a, \mv{\theta}_B) \right\}, 
 \end{equation}  
 where $\mv{\theta}_B$ are the random-effects parameters. 
 The relationships are not available in closed-form for 
 the widely used random-effects distributions, 
 e.g. Normal distribution, due to the logit link. 
 Using Bridge 
 distributional assumptions one can obtain the analytical relationship as follows. 
 If one considers $U_i = U_i^*/\phi_V$, 
 where $[U_i^*] = \mbox{Bridge}(\phi_{U^*})$, 
 and $[V_{ij}] = \mbox{Bridge}(\phi_{V})$,  
 with $0 < \phi_{U^*}, \phi_{V} < 1$, 
 then, the marginal estimates are available analytically 
 as $\alpha^m_a = \phi_{U^*} \phi_{V} \alpha^c_a$ 
 and $\beta^m_a = \phi_{U^*} \phi_{V} \beta^c_a$ \citep{wang2003, boehm2013}. 
 For the concept of bridging, one can see \cite{kenward2016}. 
 Note that the distribution of $U_i$ is no longer Bridge; we call it Modified Bridge. 
 Density functions of the Bridge and Modified Bridge are given in Appendix \ref{sec:appendix_dens}. 
 Both distributions are symmetric and zero-mean. Under Bridge, $V_{ij}$ is zero-mean, 
 and has a variance of $\frac{\pi^2}{3} ({\phi_V}^{-2} - 1)$. 
 Under Modified Bridge, $U_{i}$ is zero-mean and has a variance of $\frac{\pi^2}{3 \phi_V2} ({\phi_{U^*}}^{-2} - 1)$. 
 Bridge density is plotted in Figure \ref{fig:distr} against Normal 
 for two settings of variance. 

 We also consider Normal distribution for the random-effects: 
 $[U_i|\sigma_U] = \mbox{N}(0, \sigma_U^2)$ and $[V_{ij}|\sigma_V] = \mbox{N}(0, \sigma_V^2)$. Note that the model with Normal random-effects  corresponds to the 
 models  
 considered in \cite{raman2005} and \cite{chan2015}. 

 \subsection{Bayesian Inference}
 \label{sec:bayesian_inference} 

 Let's denote the responses in the data-set by ${\mv Y} = ({\mv Y}_1^\top, \ldots, {\mv Y}_n^\top)^\top$, where 
 ${\mv Y}_i = ({\mv Y}_{i1}^\top, \ldots, {\mv Y}_{im_i}^\top)^\top$ and ${\mv Y}_{ij} = (Y_{ij1}, \ldots, Y_{ijs_{ij}})^\top$. 
 Also let's denote the covariates by ${\mv X} = \mbox{rbind}({\mv X}_1, \ldots, {\mv X}_n)$, 
 where ${\mv X}_i = \mbox{rbind}({\mv X}_{i1}, \ldots, {\mv X}_{im_i})$, 
 ${\mv X}_{ij} = \mbox{rbind}({\mv X}_{ij1}, \ldots, {\mv X}_{ijs_{ij}})$ (with ${\mv X}_{ijk}$ as before), 
 with $\mbox{rbind}(\cdot)$ being an operator that stacks the matrices on bottom of each other. 
 
 The joint posterior distribution of the parameters,
 ${\mv \theta} = ({{\mv \theta}_{{\mv \alpha}^c}}^\top, {{\mv \theta}_{{\mv \beta}^c}}^\top, {{\mv \theta}_{\mv B}}^\top)^\top$, 
 where 
 ${\mv \theta}_{{\mv \alpha}^c} = (\alpha^c_1, \ldots, \alpha^c_{A - 1})^\top$, 
 and 
 ${\mv \theta}_{{\mv \beta}^c} = ({{\mv \beta}^c_1}^\top, \ldots, {{\mv \beta}^c_{A - 1}}^\top)^\top$,  
 and random effects, $\mv B$, where $\mv B = ({\mv B}_1^\top, \ldots, {\mv B}_n^\top)^\top$, where 
 ${\mv B}_i = ({\mv B}_{i1}^\top, \ldots, {\mv B}_{im_i}^\top)^\top$ and ${\mv B}_{ij} = (B_{ij1}, \ldots, B_{ijs_{ij}})^\top$
 given data, $\mv X$ and $\mv Y$, is given by
 \begin{align}	
 [\mv \theta, \mv B | \mv Y, \mv X ] & \propto [\mv Y | \mv \theta, \mv B, \mv X] [\mv B | \mv \theta, \mv X] [\mv \theta, \mv X], \nonumber \\
 \label{eq:joint_posterior}
 & \propto [\mv Y | {\mv \theta}_{{\mv \alpha}^c}, {\mv \theta}_{{\mv \beta}^c}, \mv B, \mv X]  [\mv B | {\mv \theta}_{\mv B}, \mv X] [{\mv \theta}_{{\mv \alpha}^c}][{\mv \theta}_{{\mv \beta}^c}] [{\mv \theta}_{\mv B}| \mv X].
 \end{align}
 By assuming random-effects and associated parameters are independent of the covariates, the joint posterior 
 \eqref{eq:joint_posterior} can be re-written as
 \begin{align}	
\label{eq:joint_posterior2}
 [\mv \theta, \mv B | \mv Y, \mv X ] & \propto [\mv Y | {\mv \theta}_{{\mv \alpha}^c}, {\mv \theta}_{{\mv \beta}^c}, \mv B, \mv X] [\mv B | \mv \theta_B] [{\mv \theta}_{{\mv \alpha}^c}][{\mv \theta}_{{\mv \beta}^c}] [{\mv \theta}_{\mv B}]. 
 \end{align}
 Here, $[\mv Y | {\mv \theta}_{{\mv \alpha}^c}, {\mv \theta}_{{\mv \beta}^c}, \mv B, \mv X]$ is the likelihood: 
 \begin{align}
 \label{eq:likelihood}
 \prod_{i=1}^{n} \prod_{j=1}^{m_i} \prod_{k=1}^{s_{ij}} [Y_{ijk}|{\mv \theta}_{{\mv \alpha}^c}, {\mv \theta}_{{\mv \beta}^c}, B_{ijk}, \mv X_{ijk}].
 \end{align}
 In \eqref{eq:likelihood}, $[Y_{ijk}|{\mv \theta}_{{\mv \alpha}^c}, {\mv \theta}_{{\mv \beta}^c}, B_{ijk}, \mv X_{ijk}]$ 
 is multinomial distribution such that 
 \begin{align}
 [Y_{ijk}|{\mv \theta}_{{\mv \alpha}^c}, {\mv \theta}_{{\mv \beta}^c}, B_{ijk}, \mv X_{ijk}] = 
 \prod_{a = 1}^{A} \mbox{P}(Y_{ijk} = a | {\mv \theta}_{{\mv \alpha}^c}, {\mv \theta}_{{\mv \beta}^c}, B_{ijk}, \mv X_{ijk}),
 \end{align}
 with $\mbox{P}(Y_{ijk} = a | {\mv \theta}_{{\mv \alpha}^c}, {\mv \theta}_{{\mv \beta}^c}, B_{ijk}, \mv X_{ijk})$ to be obtained 
 from \eqref{eq:glmm_cumlogit}. 
 
 By specifying $B_{ijk} = U_i + V_{ij}$, with $U_i = U_i^*/e$, one would obtain 
 $[\mv B | {\mv \theta}_{\mv B}]$ as  
 \begin{align}
 \prod_{i = 1}^{n} [U_i^* | {\mv \theta}_{U^*}] \prod_{i = 1}^{n} \prod_{j = 1}^{m_i} [V_{ij} | {\mv \theta}_V].
 \end{align}
 For Bridge random-effects, 
 $[U_i^* | {\mv \theta}_{U^*}] = \mbox{Bridge}(U_i^* | {\mv \theta}_{U^*})$,
 $[V_{ij} | {\mv \theta}_V] = \mbox{Bridge}(V_{ij} | {\mv \theta}_V)$, 
 ${\mv \theta}_{U^*} = \phi_{U^*}$, ${\mv \theta}_{V} = \phi_{V}$, and $e = \phi_{V}$. 
 For Normal, 
 $[U_i^* | {\mv \theta}_{U^*}] = \mbox{N}(0, \sigma_{U^*}^2)$,
 $[V_{ij} | {\mv \theta}_V] = \mbox{N}(0, \sigma_{V}^2)$, 
 ${\mv \theta}_{U^*} = \sigma_{U^*}$, ${\mv \theta}_{V} = \sigma_{V}$. 
 Note that for Normal, $e = 1$. 

 Weakly informative priors are specified for the parameters.   
 For ${\mv \theta}_{{\mv \alpha}^c}$ and ${\mv \theta}_{{\mv \beta}^c}$, 
 we consider Cauchy priors with location parameter 0 and 
 scale parameter 5 \citep{gelman2008}. 
 Standard deviations of the distributions, e.g. for $\sqrt{\frac{\pi^2}{3} (\phi^{-2} - 1)}$ for Bridge,
 are assigned half-Cauchy with location 0 and scale 5 \citep{gelman2006, polson2012}.   
 
 Samples from the joint posterior density are drawn using 
 The No-U-Turn Sampler of \cite{hoffman2014}, a modified version 
 of Hamiltonian Monte Carlo \citep{neal2011},
 as implemented in {\tt Stan} \citep{carpenter2017}. 
 Computations were 
 carried out in {\tt R} \citep{r2018} using bespoke code that relies on 
 the {\tt rstan} package \citep{rstan2018}. 
 Bespoke R codes, a simulated data-set, and 
 exemplary codes for data analysis are available in the {\tt R} package 
 {\tt mixed3} (\url{https://github.com/ozgurasarstat/mixed3}). 
 
 \section{Application: TR-SILC data-set}
 \label{sec:silc_analysis}

 \subsection{Results}
 \label{sec:results}

 As mentioned in Section \ref{sec:data}, the outcome variable is re-categorised SRH with 
 three levels ($A = 3$): 
 good health ($Y_{ijk} = 1$), fair health ($Y_{ijk} = 2$), and poor health ($Y_{ijk} = 3$). 
 Explanatory variables are listed in Tables \ref{tab:summary_stats1} \& \ref{tab:summary_stats2}. 

 We fit to the TR-SILC data-set the three-level models
 with Modified Bridge distributed $U_i$ and 
 Bridge distributed $V_{ij}$ (Modified Bridge - Bridge model), and 
 Normally distributed $U_i$ and Normally distributed $V_{ij}$ (Normal - Normal model).  
 We also fit the two-level mixed model with  
 Bridge distributed $V_{ij}$ by dropping $U_i$ from the model (two-level Bridge model), 
 and the fixed-effects  model by dropping both $U_i$ and $V_{ij}$.
 For all the four models, we specifically consider proportional odds assumption, 
 i.e. ${\mv \beta}_a^c = {\mv \beta}^c$, as there are fairly 
 many explanatory variables. For each model, four chains of length 2,000 
 were run in parallel starting from random initials. 
  First halves of the chains were treated as the warm-up period. 
 Trace-plots, density plots and R-hat statistics \citep{brooks97} were used to assess convergence. 
 Fitting each of the three-level models 
 took around ten hours, 
 whereas it took around nine hours 
 for the two-level model and around an hour the fixed-effects model, 
 on a 64 bit desktop computer with 
 16,00 GB RAM and AMD Ryzen 7 1800X Eight Core Processor 
 3.60 GHz running Windows 10. 
 
  We consider the  
 Watanebe information criterion (WIC; \cite{watanebe2010}, \cite{gelman2014}) and log  
 pseudo marginal likelihood (LPML; \cite{dey1997}, 
 \cite{gelman2014}) 
 for model comparison. 
 The formulae that we used to calculate these criteria 
 are given in Appendix \ref{sec:appendix_model_selection}. 
 Lower values of WIC and higher values of  
 LPML indicate better fit. 
 Table \ref{tab:criteria} displays the values obtained 
 for the models used to analyse the TR-SILC data-set. 
 Both WAIC and LPML indicate that the Modified Bridge - Bridge 
 model is the best fitting model. Normal - Normal model is 
 the second, two-level Bridge is the third and 
 the fixed-effects model is the fourth model. 
  
  \begin{table}[t]
 \begin{center}
 \caption{Model selection criteria. }
 \label{tab:criteria}
 \fbox{
 \scalebox{1}{
 \begin{tabular}{l c c} 
 Model                    & WAIC      & LPML \\ \hline 
 Modified Bridge - Bridge & 130 890.4 & -66 451.4 \\
 Normal - Normal          & 131 429.5 & -66 505.5 \\
 Two-level Bridge         & 132 273.1 & -67 291.8 \\
 Fixed                    & 152 981.8 & -76 490.9 \\
 \end{tabular}}}
 \end{center}
 \end{table}  
 
 Conditional results are presented in 
 Table \ref{tab:results_conditional}, whereas marginal results 
 are presented in Table \ref{tab:results_marginal}. 
 With regards the conditional results, ${\mv \beta}^c$, 
 different random-effects 
 distributional assumptions for three-level models 
 produced similar results. This would be 
 expected, since all the 
 distributions are zero-mean and symmetric. 
 In addition, 
 this would be considered as a good sign, 
 because Modified Bridge - Bridge pair 
 can be used instead of the widely used Normal 
 distribution, and the pair brings the advantage 
 of direct marginal inferences as 
 discussed in Section \ref{sec:ref_distr}. 
 The two-level Bridge model also 
 produced similar results up to some extent. 
 With regards marginal results, 
 ${\mv \beta}^m$, 
 Modified Bridge - Bridge and two-level Bridge 
 models again produced similar results. 
 There are considerable differences between these two models and 
 the fixed-effects model. For example, for ``Students", whilst 
 the 95\% credibility interval is (-0.242, -0.035) 
 under the Modified Bridge - Bridge model, 
 it is (-0.418, -0.211) under the fixed-effects model. 
 
 Interpretations under the preferred model (Modified Bridge - Bridge model) 
 are as follows. 
 Females were approximately 29\% $(= (\exp(0.251)-1)*100)$ more 
 likely to report worse health compared to males. 
 Widowed or separated (never married) people were more (less) likely to report worse 
 health compared to married people.  
 People in the 35 -- 64 (65+) age group were less (more) likely to report worse health 
 compared to those in the 16 -- 34 age group. Higher education level was associated with 
 decreased probability of reporting worse health. 
 Except students, all the other working status categories were more 
 likely to report worse health compared to those working full or part time. 
 Unemployed people were approximately 22\% more likely to report worse health 
 compared to the full or part time working people. MHDI  
 is inversely associated with SRH such that 1\% decrease in MHDI level was associated with 
 approximately 32\% increased odds of reporting worse health. Odds of 
 reporting worse health did not change in year 2011 compared to 2010; the credibility 
 interval for $\beta_{14}^m$ is (-0.082, 0.011). On the other hand, 
 in 2012 and 2013 people were less likely to report worse health compared to 2010.   
 They were approximately 12\% (9\%) more likely to report worse health in 2010 compared to 2012 (2013). 
  
 \subsection{Posterior predictive checks}  
 \label{sec:select_ppc}  
  
  We performed posterior predictive checks to see 
  how the replicated data based on the fitted models 
  match the observed data. For this, we simulated 
  data from 
  \begin{align}
  [Y_{ijk}^{rep} | \mv{Y}] = \int \int [Y_{ijk}^{rep} | B_{ijk}, \mv{\theta}] [B_{ijk} | \mv{Y}, \mv{\theta}] [\mv{\theta}|\mv{Y}] d B_{ijk} d\mv{\theta} 
  \end{align}
 for the Modified Bridge - Bridge, Normal - Normal and the 
 two-level Bridge models, and from   
  \begin{align}
  [Y_{ijk}^{rep} | \mv{Y}] = \int [Y_{ijk}^{rep} | \mv{\theta}] [\mv{\theta}|\mv{Y}] d\mv{\theta} 
  \end{align}  
 for the fixed-effects model, 
 for each of the 4 000 elements of the MCMC samples. 
 Each of the 4 000 simulated data-sets were then compared 
 with the observed data. We calculated percentages of 
 matches and mis-matches between the observed and replicated data-sets.  
 Means and standard deviations of the percentages  
 are displayed in Table \ref{tab:ppc}. 
 In this table, 
 ``-2" means observed 
 outcome being ``good health" and replicated being ``poor health"; 
 ``-1" means observed being ``good health" and replicated being ``fair health", 
 or observed being ``fair health" and replicated being ``poor health"; 
 ``0" means observed and replicated being the same; 
 ``1" means observed being ``fair health" and replicated being ``good health", 
 or observed being ``poor health" and replicated being ``fair health"; 
 ``2" means observed being ``poor health" and replicated being ``good health".
 In summary, non-zero values mean mis-match between the observed 
 and replicated data-sets, whilst 
 ``-2" and ``2" would mean the 
 most difference between the two. 
 Modified Bridge - Bridge, Normal-Normal and two-level Bridge models 
 seem to be perform similarly in terms of 
 replicating the observed data. Fixed-effects model 
 seems to be the worst amongst the four models, as 
 its mean percentage for ``0" is lower and 
 mean percentages 
 for ``-2" and ``2" are higher compared to the other models.  
 
 \begin{table}[t]
 \begin{center}
 \caption{Posterior predictive checks. For details, see the text. }
 \label{tab:ppc}
 \fbox{
 \scalebox{0.8}{
 \begin{tabular}{l c c c c c c c c c c} 
                    & \multicolumn{2}{c}{-2} & \multicolumn{2}{c}{-1} & \multicolumn{2}{c}{0} & \multicolumn{2}{c}{1} & \multicolumn{2}{c}{2} \\ \hline 
  & Mean & SD & Mean & SD & Mean & SD & Mean & SD & Mean & SD \\ \cmidrule(lr){2-11}
 Modified Bridge - Bridge & 4.27 & 0.07 & 13.53 & 0.13 & 57.64 & 0.13 & 17.05 & 0.08 & 7.52 & 0.05 \\
 Normal - Normal  & 4.55 & 0.08 & 13.19 & 0.12 & 57.58 & 0.12 & 17.01 & 0.07 & 7.67 & 0.05 \\
 Two-level Bridge & 4.36 & 0.07 & 13.85 & 0.13 & 57.56 & 0.13 & 16.89 & 0.08 & 7.35 & 0.05 \\ 
 Fixed-effects & 6.71 & 0.08 & 15.70 & 0.12 & 55.05 & 0.13 & 15.80 & 0.08 & 6.74 & 0.05 \\ 
 \end{tabular}}}
 \end{center}
 \end{table}

 \section{Conclusion and Discussion}
 \label{sec:conclusion}
 
 In this paper, we presented an analysis of Turkish 
 Income and Living Conditions data, 
 with the perspective of marginal inference on the relationships between the 
 outcome and explanatory variables.  
 The outcome of interest is ordinal taking three values: good, fair and bad health. It is subject to family- and 
 individual-level dependencies. In other words, it is three-level or clustered longitudinal. 
 The model we consider is cumulative logistic regression. 
 We introduced random-effects in order for likelihood-based inference. 
 Typically the estimates obtained under the random-effects models have conditional interpretations, 
 hence do not address the scientific interests of the current work. Bridge distributional 
 assumptions for the random-effects allows one to obtain marginal inferences analytically. 
 We take a Bayesian perspective for inference. Samples from the joint posterior distributions 
 are drawn using an adaptive Hamiltonian Monte Carlo algorithm, called the No-U-Turn Sampler. 
  The {\tt R} package {\tt mixed3} contains functions 
  that implement the methods in practice.  	

 

 The findings regarding working status, 
 MHDI and panel years 
 are important from the public health point of view. 
 The difference between employed and unemployed people 
 in terms of reporting better health 
 emphasises the importance of reducing unemployment rate. 
 Inverse association between MDHI levels and probability 
 of reporting better health points out the vitality of 
 economic conditions to reaching better healthcare.  
 Differences in cohort years are important to understand 
 the impacts of changes in health policies.
		

 In the panels of TR-SILC, neither geographical nor urban/rural information 
 was available. These information would explain some source of variation in SRH. 
 Mediation analysis would be interesting. 
 For example, education level would 
 effect income and hence health status. 
 Cumulative and/or lagged effects of 
 explanatory variables might be considered to  
 explain health status. For example, 
 history of unemployment might be predictor of current health status. 
 We leave these to a future work on substantive analysis of TR-SILC. 
 Marginalised models \citep{lee2007} could be extended to three-level 
 ordinal outcomes. However, the estimation procedure would be 
 computationally more demanding compared to the present approach, 
 as one needs to solve intractable convolution equations through root finding 
 algorithms, e.g. Newton-Raphson.  
 Currently we do have access only to TR-SILC. It would be interesting if 
 we were able to add EU and other SILC data into our analysis. 
 In such a case, the data would be four-level due to the additional nested 
 structure of families being nested within countries. It would be interesting 
 to explore Bridge random-effects specification for four-level ordinal outcomes. 
 From practical point of view, however, one can include countries as dummy variables 
 in the design matrix, as there will be 32 countries.  


\begin{table}
\begin{center}
\caption{Conditional results for TR-SILC data. 
Two-level Bridge is the model with 
Bridge $V_{ij}$. 
``CI" stands for credibility interval.}
\label{tab:results_conditional}
\fbox{
\scalebox{0.60}{
\begin{tabular}{l l r r r r r r r r r r } 
& & \multicolumn{3}{c}{Modified Bridge - Bridge} & \multicolumn{3}{c}{Normal - Normal} & \multicolumn{3}{c}{Two-level Bridge}\\ \cmidrule(lr){3-11}
Variable               & \ \ \     & Mean         & SD    & 95\% CI        & \ \ \ Mean  & SD    & 95\% CI       & Mean         & SD    & 95\% CI \\ \hline
Threshold     & \ \ \ $\alpha_1^c$ & \ \ \ -1.871 & 0.206 & -2.277, -1.473 &\ \ \ -1.697 & 0.206 &-2.093, -1.283 & \ \ \ -1.914 & 0.191 & -2.299, -1.548 \\ 
Threshold     & \ \ \ $\alpha_2^c$ & \ \ \ 0.520  & 0.206 & 0.117, 0.920   & \ \ \  0.685& 0.207 & 0.288, 1.095  & \ \ \ 0.471  & 0.191 & 0.086,   0.834 \\ 
Male (Ref)    & \ \ \              & \ \ \        &       &                & \ \ \       &       &               & \ \ \        &       &                \\
Female        & \ \ \ $\beta_1^c$  & \ \ \ 0.383  & 0.031 & 0.323, 0.443   & \ \ \ 0.348 & 0.032 & 0.286, 0.412  & \ \ \ 0.385  & 0.034 & 0.321, 0.451   \\  
Married (Ref) & \ \ \              & \ \ \        &       &                & \ \ \       &       &               & \ \ \        &       &                \\ 
Never married & \ \ \ $\beta_2^c$  & \ \ \ -0.549 & 0.051 & -0.650, -0.449 & \ \ \ -0.481& 0.048 &-0.576, -0.386 & \ \ \ -0.542 & 0.050 & -0.642, -0.445 \\  
Widowed/separated & \ \ \ $\beta_3^c$  & \ \ \ 0.496  & 0.044 & 0.413, 0.582   & \ \ \ 0.515 & 0.046 & 0.426, 0.603  & \ \ \ 0.506  & 0.045 & 0.418, 0.593 \\ 
Age (16--34) (Ref) & \ \ \               & \ \ \         &       &                & \ \ \       &       &               & \ \ \        &       & \\ 
Age (35--64) & \ \ \ $\beta_4^c$   & \ \ \ -1.881  & 0.038 & -1.957, -1.809 & \ \ \ -1.831& 0.037 &-1.904, -1.759 &\ \ \ -1.815  & 0.038 & -1.890, -1.739\\
Age (65+)    & \ \ \ $\beta_5^c$  & \ \ \ 1.277  & 0.040 & 1.198, 1.355   & \ \ \ 1.297 & 0.041 & 1.216, 1.376  & \ \ \ 1.314  & 0.040 & 1.238, 1.393\\ 
Higher education (Ref) & \ \ \              & \ \ \        &       &                & \ \ \       &       &               & \ \ \        &       & \\ 
Primary/less    & \ \ \ $\beta_6^c$  & \ \ \ 1.412  & 0.060 & 1.293, 1.529   & \ \ \ 1.446 & 0.061 & 1.331, 1.569  & \ \ \ 1.458  & 0.060 & 1.341, 1.575\\ 
Secondary/high  & \ \ \ $\beta_7^c$  & \ \ \ 0.441  & 0.060 & 0.320, 0.556   & \ \ \ 0.441 & 0.059 & 0.325, 0.559  & \ \ \ 0.427  & 0.061 & 0.308, 0.546\\ 
Full - part time (Ref) & \ \ \              & \ \ \        &       &                & \ \ \       &       &               & \ \ \        &       & \\ 
Housekeeper   & \ \ \ $\beta_8^c$  & \ \ \ 0.339  & 0.033 & 0.273, 0.405   & \ \ \ 0.354 & 0.035 & 0.286, 0.421  & \ \ \ 0.316  & 0.035 & 0.248, 0.384\\  
Retired      & \ \ \ $\beta_9^c$    & \ \ \ 0.855  & 0.042 & 0.771, 0.936   & \ \ \ 0.865 & 0.044 & 0.778, 0.953  & \ \ \ 0.915  & 0.044 & 0.826, 0.999\\ 
Student      & \ \ \ $\beta_{10}^c$ & \ \ \ -0.212 & 0.081 & -0.370, -0.053 &\ \ \ -0.199 & 0.074 &-0.346, -0.055 & \ \ \ -0.368 & 0.083 & -0.532, -0.210\\  
Unemployed   & \ \ \ $\beta_{11}^c$ & \ \ \ 0.302  & 0.060 & 0.187, 0.422   & \ \ \ 0.292 & 0.060 & 0.176, 0.407  & \ \ \ 0.301  & 0.060 & 0.182, 0.419\\  
Others       & \ \ \ $\beta_{12}^c$ & \ \ \ 2.431  & 0.046 & 2.341, 2.521   & \ \ \ 2.488 & 0.047 & 2.396, 2.581  & \ \ \ 2.374  & 0.047 & 2.282, 2.466\\  
$\log(\mbox{MHDI} + 1)$& \ \ \ $\beta_{13}^c$ & \ \ \ -0.425 & 0.021 & -0.466, -0.385 &\ \ \ -0.413 & 0.021 &-0.454, -0.371 & \ \ \ -0.438 & 0.019 & -0.475, -0.403\\  
Year = 2010 (Ref)     & \ \ \                & \ \ \        &       &                & \ \ \       &       &               & \ \ \        &       & \\  
Year = 2011   & \ \ \ $\beta_{14}^c$ & \ \ \ -0.056 & 0.037 & -0.125, 0.017  & \ \ \ -0.058& 0.036 &-0.129, 0.013  & \ \ \ -0.052 & 0.036 & -0.124, 0.018\\  
Year = 2012  & \ \ \ $\beta_{15}^c$ & \ \ \ -0.177 & 0.036 & -0.245, -0.107 & \ \ \ -0.177& 0.036 &-0.249, -0.107 & \ \ \ -0.168 & 0.035 & -0.235, -0.101\\  
Year = 2013  & \ \ \ $\beta_{16}^c$ & \ \ \ -0.136 & 0.036 & -0.205, -0.065 & \ \ \ -0.136& 0.036 &-0.209, -0.066 & \ \ \ -0.127 & 0.035 &-0.195, -0.060 \\ 
             & \ \ \ $\phi_{U^*}$   & \ \ \ 0.865  & 0.004 & 0.856, 0.873   & \ \ \       &       &               & \ \ \        &       & \\
             & \ \ \ $\phi_V$       & \ \ \ 0.757  & 0.004 & 0.748, 0.765   & \ \ \       &       &               & \ \ \ 0.667  & 0.004 & 0.659, 0.674 \\ 
             & \ \ \ $\sigma_U$     & \ \ \        &       &                & \ \ \ 1.175 & 0.020 & 1.135, 1.216  & \ \ \        &       & \\
			 & \ \ \ $\sigma_V$     & \ \ \        &       &                & \ \ \ 1.473 & 0.020 & 1.434, 1.514  & \ \ \        &       & \\
\end{tabular}}}
\end{center}
\end{table}


\begin{table}[htbp]
\begin{center}
\caption{Marginal results for TR-SILC data. 
Note that marginal results for threshold parameters are not reported. }
\label{tab:results_marginal}
\fbox{
\scalebox{0.6}{
\begin{tabular}{l l r r r r r r r r r } 
                       &                & \multicolumn{3}{c}{Modified Bridge - Bridge} & \multicolumn{3}{c}{Two-level Bridge}  & \multicolumn{3}{c}{Fixed-effects} \\ 
											\cmidrule(lr){3-11} 
Variable               &                & Mean         & SD    & 95\% CI       & Mean   & SD    & 95\% CI       & Mean  & SD   & 95\% CI  \\ \hline
Male (Ref)             &                & \ \ \        &       &               & \ \ \         &       &               &  \ \ \ 	  &      & \\
Female                 & $\beta_1^m$    & \ \ \ 0.251  & 0.020 & 0.212, 0.290  & \ \ \  0.257  & 0.022 & 0.214, 0.301  &  \ \ \  0.296 & 0.019& 0.259, 0.333\\  
Married (Ref)          &                & \ \ \        &       &               & \ \ \         & 	  & 			  & \ \ \     	  &      &  \\ 
Never married          & $\beta_2^m$    & \ \ \ -0.360 & 0.033 &-0.426, -0.294 & \ \ \  -0.361 & 0.033 &-0.428, -0.297 & \ \ \  -0.286& 0.028& -0.340, -0.232\\  
Widowed or separated   & $\beta_3^m$    & \ \ \ 0.325  & 0.029 & 0.270, 0.381  & \ \ \  0.338  & 0.030 & 0.279, 0.396  & \ \ \  0.268 & 0.025& 0.219, 0.317\\ 
Age (16--34) (Ref)     &                & \ \ \        &       & 			   & \ \ \  		  & 	  & 		      & \ \ \     	  &      & \\ 
Age (35--64)           & $\beta_4^m$    & \ \ \ -1.231 & 0.025 &-1.280, -1.184 & \ \ \  -1.210 & 0.025 & -1.259, -1.160& \ \ \  -1.306& 0.021& -1.349, -1.264\\
Age (65+)              & $\beta_5^m$    & \ \ \ 0.836  & 0.026 & 0.784, 0.886  & \ \ \  0.876  & 0.027 & 0.825, 0.927  & \ \ \  0.830 & 0.022& 0.787, 0.873\\ 
Higher education (Ref) &                & \ \ \        &       &  			   & \ \ \  		  & 	  & 		   	  & \ \ \  	  &      & \\ 
Primary or less        & $\beta_6^m$    & \ \ \ 0.924  & 0.039 & 0.846, 1.001  & \ \ \  0.972  & 0.039 & 0.894, 1.049  & \ \ \  0.885 & 0.033& 0.820, 0.947\\ 
Secondary or high      & $\beta_7^m$    & \ \ \ 0.289  & 0.040 & 0.209, 0.363  & \ \ \  0.285  & 0.041 &  0.204, 0.363 & \ \ \  0.241 & 0.033& 0.177, 0.308\\ 
Full - part time (Ref) &                & \ \ \        &       & 			   & \ \ \  		  & 	  &   			  & \ \ \     	  &      & \\
Housekeeper            & $\beta_8^m$    & \ \ \ 0.222  & 0.022 & 0.179, 0.265  & \ \ \  0.211  & 0.023 &  0.165, 0.256 & \ \ \  0.230 & 0.020& 0.190, 0.271\\  
Retired                & $\beta_9^m$    & \ \ \ 0.559  & 0.028 & 0.504, 0.613  & \ \ \  0.610  & 0.030 & 0.550, 0.666  & \ \ \  0.733 & 0.025& 0.684, 0.781\\ 
Student                & $\beta_{10}^m$ & \ \ \ -0.139 & 0.053 &-0.242, -0.035 & \ \ \  -0.245 & 0.055 & -0.355, -0.140 & \ \ \  -0.315& 0.053& -0.418, -0.211\\  
Unemployed             & $\beta_{11}^m$ & \ \ \ 0.198  & 0.039 & 0.122, 0.276  & \ \ \  0.201  & 0.040 &  0.121, 0.279 & \ \ \  0.200 & 0.042& 0.118, 0.282\\  
Others                 & $\beta_{12}^m$ & \ \ \ 1.591  & 0.031 & 1.532, 1.651  & \ \ \  1.583  & 0.032 & 1.520, 1.646  & \ \ \  2.016 & 0.029& 1.958, 2.073\\  
$\log(MHDI + 1)$       & $\beta_{13}^m$ & \ \ \ -0.278 & 0.014 &-0.305, -0.252 & \ \ \  -0.292 & 0.013 &-0.318, -0.268 & \ \ \  -0.355& 0.011& -0.377, -0.334\\  
Year = 2010 (Ref)      &                & \ \ \        &       & 			   & \ \ \         & 	  & 	   		  & \ \ \        &      & \\  
Year = 2011            & $\beta_{14}^m$ & \ \ \ -0.037 & 0.024 & -0.082, 0.011 & \ \ \  -0.035 & 0.024 & -0.083, 0.012 & \ \ \  -0.020& 0.029& -0.078, 0.037\\  
Year = 2012            & $\beta_{15}^m$ & \ \ \ -0.116 & 0.023 & -0.161, -0.070& \ \ \  -0.112 & 0.023 & -0.157, -0.067& \ \ \  -0.092& 0.028& -0.147, -0.038\\  
Year = 2013            & $\beta_{16}^m$ & \ \ \ -0.089 & 0.024 & -0.134, -0.043& \ \ \  -0.084 & 0.023 & -0.130, -0.040& \ \ \  -0.057& 0.028& -0.113, -0.003\\
\end{tabular}}}
\end{center}
\end{table}

%
%

\section*{Acknowledgements}
 The author thanks to Dr. Mahmut Yard{\i}m (Hacettepe University)
 for bringing the TR-SILC data-set into his attention and helpful discussions, 
 Prof. Peter Diggle (Lancaster University) for helpful comments on a draft, 
 Dr. Jonas Wallin (Lund University) and 
 Dr. David Bolin (University of Gothenburg and Chalmers University of Technology) 
 for helpful discussions on modelling and inference, 
 and TURKSTAT for making the data available. 
 The views expressed in this paper are those of the author.

 \appendix

 \section{Density functions}  
 \label{sec:appendix_dens}
	
	The density function of Bridge 
 distribution for logit link is given by
 \begin{align}
 \label{eq:bridge_dist}
 f(x | \phi)= \frac{1}{2 \pi} \frac{sin(\phi \pi)}{cosh(\phi x) + cos(\phi \pi)}, \ \ -\infty < x < \infty, \ 0 < \phi < 1,  
 \end{align}
 where $cosh(\cdot)$ is the hyperbolic cosine, defined as $cosh(x) = \frac{1}{2} (\exp(x) + \exp(-x))$. 
 Bridge distribution is symmetric, zero-mean and has a variance of $\frac{\pi^2}{3} (\phi^{-2} - 1)$. 
 The density function of modified Bridge, 
 for generic $X$, $Y$ and $Z$ with $X = Y/{\phi_Z}$, $[Y|\phi_Y] = \mbox{Bridge}(\phi_Y)$, $[Z|\phi_Z] = \mbox{Bridge}(\phi_Z)$, 
 is given by
 \begin{align}
 \label{eq:mod_bridge_dist}
 f(x | \phi_Y, \phi_Z)= \frac{\phi_Z}{2 \pi} \frac{sin(\phi_Y \pi)}{cosh(\phi_Y \phi_Z x) + cos(\phi_Y \pi)}, 
 \ \ -\infty < x < \infty, \ 0 < \phi_Y, \phi_Z < 1,  
 \end{align}
 Modified Bridge is zero-mean, and has a variance of $\frac{\pi^2}{3 \phi_Z^2} (\phi_Y^{-2} - 1)$. 
 	
 \section{Model selection}
 \label{sec:appendix_model_selection}	
 
 WAIC is calculated as 
 \begin{align}
 \mbox{WAIC} = -2(\mbox{lppd} - \rho),
 \end{align}	
 where, ``lppd" stands for log pointwise posterior density 
 and is calculated as
 \begin{align}
 \label{eq:lppd}
 \mbox{lppd} = \sum_{i = 1}^{n} \sum_{j = 1}^{m_i} \sum_{k = 1}^{s_{ij}} \log\left( \frac{1}{M} \sum_{l = 1}^{M} [Y_{ijk}|B_{ijk}^{(l)}, \mv{\theta}^{(l)}] \right),
 \end{align} 
 and $\rho$ is the effective number of parameters 
 and calculated as 
 \begin{align}
 \label{eq:rho}
 \rho = \sum_{i = 1}^{n} \sum_{j = 1}^{m_i} \sum_{k = 1}^{s_{ij}} V_{l = 1}^{M}\left( \log\left( [Y_{ijk}|B_{ijk}^{(l)}, \mv{\theta}^{(l)}] \right) \right),
 \end{align} 
 with 
 \begin{align}
 \label{eq:rho_var}
 V_{l = 1}^{M}(a) = \frac{1}{M - 1} \sum_{l = 1}^{M} (a^{(l)} - \bar{a})^2.
 \end{align}
 In \eqref{eq:lppd} - \eqref{eq:rho_var}, $B_{ijk}^{(l)}$,
 and $\mv{\theta}^{(l)}$ denote 
 $l$th draw of $B_{ijk}$ and $\mv{\theta}$, 
 from the joint posterior densities. 
 
 Conditional predictive ordinate (CPO) is defined as leave-one-out 
 cross-validated predictive density, such that 
 $CPO_{ijk} = [Y_{ijk}|\mv{Y}_{-({ijk})}]$, 
 where $\mv{Y}_{-({ijk})}$ is the $\mv{Y}$ 
 without observation $(ijk)$. 
 The harmonic mean estimate of CPO, as proposed by \cite{dey1997}, 
 is 
 \begin{align}
 \widehat{\mbox{CPO}_{ijk}} = \left\{ \frac{1}{M} \sum_{l=1}^{M} \frac{1}{[Y_{ijk}|B_{ijk}, \mv{\theta}_l]} \right\}^{-1}.
 \end{align}
 LPML is then defined as 
 \begin{align}
 \mbox{LPML} = \sum_{i = 1}^{n} \sum_{j = 1}^{m_i} \sum_{k = 1}^{s_{ij}} \log\left( \widehat{\mbox{CPO}_{ijk}} \right).
 \end{align}


\begin{thebibliography}{99}

\bibitem[Abebe et~al.(2016)]{abebe2016}
 Abebe, D. S., T{\o}ge, A. G., and Dahl, E. (2016). 
 Individual-level changes in self-rated health before and during the economic crisis in Europe. 
 \emph{International Journal for Equity in Health}, \textbf{15(1)}, 1--8.

 \bibitem[Arora et~al.(2015)]{arora2015}
 Arora, V. S., Karanikolos, M., Clair, A., Reeves, A., Stuckler, D., and McKee, M. (2015). 
 Data Resource Profile: The European Union Statistics on Income and Living Conditions (EU-SILC).
 \emph{International Journal of Epidemiology}, \textbf{44(2)}, 451--461.

 \bibitem[Bacci et~al.(2017)]{bacci2017}
 Bacci, S., Pigini, C., Seracini, M., and Minelli, L. (2017). 
 Employment condition, economic deprivation and self-evaluated health in Europe: Evidence from EU-SILC 2009–2012.
 \emph{International Journal of Environmental Research and Public Health}, \textbf{14:143}, doi.org/10.3390/ijerph14020143.

 \bibitem[Barlow et~al.(2015)]{barlow2015}
 Barlow, P., Reeves, A., McKee, M., and Stuckler, D. (2015). 
 Austerity, precariousness, and the health status of Greek labour market participants: 
 Retrospective cohort analysis of employed and unemployed persons in 2008–2009 and 2010–2011.
 \emph{Journal of Public Health Policy}, \textbf{36(4)}, 452--468.

 \bibitem[Boehm et~al.(2013)]{boehm2013}
 Boehm, L., Reich, B. J., and Bandyopadhyay, D. (2013). 
 Bridging Conditional and Marginal Inference for Spatially Referenced Binary Data. 
 \emph{Biometrics}, \textbf{69}, 545--554.

 \bibitem[Bandyopadhyay et~al.(2010)]{bandyopadhyay2010}
 Bandyopadhyay, D., Sinha, D., Lipsitz, S., and Letourneau, E. (2010). 
 Changing approaches of prosecutors towards juvenile repeated sex-offenders: A Bayesian evaluation.
 \emph{The Annals of Applied Statistics}, \textbf{4(2)}, 805--829.

 \bibitem[Brooks and Gelman(1997)Brooks and Gelman]{brooks97}
Brooks, S. P., and Gelman, A. (1997). 
General methods for monitoring convergence of iterative simulations.
\emph{Journal of Computational and Graphical Statistics}, 
\textbf{7},
434--455.

 \bibitem[Bürgin and Ritschard(2015)]{burgin2015}
 Bürgin, R., and Ritschard G. (2015). 
 Tree-based varying coefficient regression for longitudinal
ordinal responses.  
 \emph{Computational Statistics and Data Analysis}, \textbf{86}, 65--80.

 \bibitem[Bürgin and Ritschard(2017)]{burgin2017}
 Bürgin, R., and Ritschard G. (2017). 
 Coefficient-wise tree-based varying coefficient
 regression with vcrpart.  
 \emph{Journal of Statistical Software}, \textbf{80(6)}, 1--33.

 \bibitem[Burstr\"{o}m and Fredlund(2001)]{burstrom2001}
 Burstr\"{o}m, B., and Fredlund, P. (2001). 
 Self rated health: Is it as good a predictor of subsequent mortality among adults in lower as 
 well as in higher social classes? 
 \emph{Journal of Epidemiology and Community Health}, \textbf{55}, 836--840.

 \bibitem[Carpenter et~al.(2017)]{carpenter2017}
 Carpenter, B., Gelman, A., Hoffman, M. D., Lee, D., Goodrich, B., Betancourt, M., Brubaker, M., 
 Guo, J., Li, P., and Riddell, A. (2017). 
 Stan: A probabilistic programming language.
 \emph{Journal of Statistical Software}, \textbf{76(1)}, 1--32.

 \bibitem[Chan et~al.(2015)]{chan2015}
 Chan, M.-T., Yu, D., and Yau, K. K. W. (2015).
 Multilevel cumulative logistic regression model with random effects: Application to British social attitudes panel survey data. 
 \emph{Computational Statistics \& Data Analysis}, \textbf{88}, 173--186.
 
 \bibitem[Chen et~al.(2014)]{chen2014}
 Chen, Z., Yi, G. Y., and Wu, C. (2014).
Marginal analysis of longitudinal ordinal data with 
misclassification in both response and covariates. 
 \emph{Biometrical Journal}, \textbf{56}, 69--85.
 
 \bibitem[Cheon et~al.(2014)]{cheon2014}
 Cheon, K., Thoma, M. E., Kong, X., and Albert P. S. (2014).
 A mixture of transition models for 
heterogeneous longitudinal ordinal data: 
with applications to longitudinal 
bacterial vaginosis data. 
 \emph{Statistics in Medicine}, \textbf{33}, 3204--3213.

 \bibitem[Clair et~al.(2016)]{clair2016}
 Clair, A., Reeves, A., Loopstra, R., McKee, M., Dorling, D., and Stuckler, D. (2016).
 The impact of the housing crisis on self-reported health in Europe: multilevel longitudinal modelling of 27 EU countries. 
 \emph{The European Journal of Public Health}, \textbf{26(5)}, 788--793.

 \bibitem[Cowles et~al.(1996)]{cowles1996}
 Cowles, M. K., Carlin, B. P., and Connett, J. E. (1996).
 Bayesian tobit modeling of longitudinal ordinal clinical 
 trial compliance data with nonignorable missingness. 
 \emph{Journal of the American Statistical Association}, \textbf{91}, 86--98.

 \bibitem[Dey et~al(1997)]{dey1997}
 Dey, D. K., Chen, M. H., and Chang, H. (1997).
Bayesian approach for nonlinear random effects models.
 \emph{Biometrics}, \textbf{53}, 1239--1252.

 \bibitem[Diggle et~al.(2002)]{diggle2002}
 Diggle, P. J., Heagerty, P. J., Liang, K.-Y., and Zeger, S. L. (2002).
 \emph{Analysis of Longitudinal Data, 2nd edition}.
 Oxford: Oxford University Press.


 \bibitem[Ekholm et~al.(2003)]{ekholm2003}
 Ekholm, A., Jokinen, J., McDonald, J. W., and Smith, P. W. F. (2003).
Joint regression and association modeling of longitudinal ordinal Data. 
 \emph{Biometrics}, \textbf{59}, 795--803.

 \bibitem[Erus et~al.(2015)]{erus2015}
 Erus, B., Yakut-Cakar, B., Cali, S., and Adam, F. (2015).
 Health policy for the poor: An exploration on the take-up of means-tested health benefits in Turkey.
 \emph{Social Science \& Medicine}, \textbf{130}, 99--106.

 \bibitem[Ferraini et~al.(2014)]{ferraini2014}
 Ferraini, T., Nelson, K., and Sj{\"o}berg, O. (2014).
 Unemployment insurance and deteriorating self-rated health in 23 European countries. 
 \emph{Journal of Epidemiology and Community Health}, \textbf{68(7)}, 657--662.
 
  \bibitem[Jacqmin-Gadda et~al.(2010)]{gadda2010}
 Jacqmin-Gadda, H., Proust-Lima, C., and Ami\'eva, H. (2010).
 Semi-parametric latent process model for
longitudinal ordinal data: Application to
cognitive decline. 
 \emph{Statistics in Medicine}, \textbf{29}, 2723--2731.

 \bibitem[Gelman(2006)]{gelman2006}
 Gelman, A. (2006).
 Prior distributions for variance parameters in hierarchical models.
 \emph{Bayesian Analysis}, \textbf{1(3)}, 515--534.

 \bibitem[Gelman et~al.(2008)]{gelman2008}
 Gelman, A., Jakulin, A., Pittau, M. G., and Su, Y.-S. (2008).
 A weakly informative default prior distribution for logistic and other regression models.
 \emph{The Annals of Applied Statistics}, \textbf{2(4)}, 1360--1383.

 \bibitem[Gelman et~al.(2014)]{gelman2014}
 Gelman, A., Hwang, J., and Vehtari, A. (2014).
 Understanding predictive information criteria for Bayesian models.
 \emph{Statistics and Computing}, \textbf{24}, 997--1016.
  
 \bibitem[Giannoni et~al.(2016)]{giannoni2016}
 Giannoni, M., Franzini, L., and Masiero, G. (2016).
 Migrant integration policies and health inequalities in Europe. 
 \emph{BMC Public Health}, \textbf{16:463}, DOI 10.1186/s12889-016-3095-9.

 \bibitem[Heagerty and Zeger(1996)]{heagerty1996}
 Heagerty, P. J., and Zeger, S. L. (1996).
 Marginal regression models for clustered ordinal measurements. 
 \emph{Journal of the American Statistical Association}, \textbf{91}, 1024--1036.


 \bibitem[Heggeb{\o}(2015)]{heggebo2015}
 Heggeb{\o}, K. (2015).
 Unemployment in Scandinavia during an economic crisis: Cross-national differences in health selection. 
 \emph{Social Science \& Medicine}, \textbf{130}, 115--124.

 \bibitem[Hessel(2016)]{hessel2016}
 Hessel, P. (2016).
 Does retirement (really) lead to worse health among European men and women across all educational levels?
 \emph{Social Science \& Medicine}, \textbf{151}, 19--26.

 \bibitem[Hoffman and Gelman(2014)]{hoffman2014}
 Hoffman, M. D., and Gelman, A. (2014).
 The No-U-Turn sampler: adaptively setting path lengths in Hamiltonian Monte Carlo.
 \emph{Journal of Machine Learning Research}, \textbf{15}, 1593--1623.

 \bibitem[Huijts et~al.(2015)]{huijts2015}
 Huijts, T., Reeves, A., McKee, M., and Stuckler, D. (2015).
 The impacts of job loss and job recovery on self-rated health: testing the mediating role of financial strain and income. 
 \emph{European Journal of Public Health}
 \textbf{25(5)}, 
 801--806.

 \bibitem[Kaciroti et~al.(2006)]{kaciroti2006}
 Kaciroti, N. A., Raghunathan, T. E., Schork, M. A., 
 Clark, N. M., and Gong, M. (2006).
 A Bayesian approach for clustered congitudinal 
 ordinal outcome with nonignorable missing data. 
 \emph{Journal of the American Statistical Association}
 \textbf{101}, 
 435--446.
 
  \bibitem[Kauermann(2000)]{kauermann2000}
 Kauermann, G. (2000).
 Modeling longitudinal data with ordinal response by varying coefficients. 
 \emph{Biometrics}
 \textbf{56}, 
 692--698.

 \bibitem[Kenward and Molenbergs(2016)]{kenward2016}
 Kenward, M. G., and Molenbergs, G. (2011).
 A taxonomy of mixing and outcome distributions based on conjugacy and bridging.  
 \emph{Communications in Statistics - Theory and Methods}, \textbf{45(7)}, 1953--1968.
 
  \bibitem[Kosorok and Chao(1996)]{kosorok1996}
 Kosorok, M. R. and Chao, W.-H. (1996).
The analysis of longitudinal ordinal response data in continuous time.  
 \emph{Journal of the American Statistical Assoctiation}, \textbf{91}, 807--817.

 \bibitem[Laffont et~al.(2014)]{laffont2014}
 Laffont, C. M., Vandemeulebroecke, M., and Concordet, D. (2014).
 Multivariate analysis of longitudinal ordinal data with 
 mixed effects models, with application to clinical 
 outcomes in osteoarthritis. 
 \emph{Journal of the American Statistical Assoctiation}, \textbf{109}, 955--966.

 \bibitem[Lee and Daniels(2007)]{lee2007}
 Lee, K. and Daniels, M. J. (2007).
 A class of Markov models for longitudinal ordinal data. 
 \emph{Biometrics}, \textbf{63}, 1060--1067.

 \bibitem[Lee and Daniels(2008)]{lee2008}
 Lee, K. and Daniels, M. J. (2008).
 Marginalized models for longitudinal ordinal data 
 with application to quality of life studies. 
 \emph{Statistics in Medicine}, \textbf{27}, 4359--4380.

 \bibitem[Lee et~al(2013)]{lee2013}
 Lee, K., Daniels, M. J. and Joo, Y. (2013).
 Flexible marginalized models for bivariate longitudinal 
 ordinal data. 
 \emph{Biostatistics}, \textbf{14}, 462--476.
 
 \bibitem[Lee et~al.(2016)]{lee2016}
 Lee, K., Sohn, I. and Kim, D. (2016).
 Analysis of long series of longitudinal ordinal 
 data using marginalized models. 
 \emph{Computational Statistics and Data Analysis}, \textbf{94}, 363--371.

 \bibitem[Lesaffre et~al.(1996)]{lesaffre1996}
 Lesaffre, E., Molenberghs, G., and Dewulf, L. (1996).
 Effect of dropouts in a longitudinal study: an application to 
 a repeated ordinal model. 
 \emph{Statistics in Medicine}, \textbf{15}, 1123--1141. 

 \bibitem[Li et~al.(2011)]{li2011}
 Li, X., Bandyopadhyay, D., Lipsitz, S., and Sinha, D. (2011).
 Likelihood Methods for Binary Responses of Present Components
in a Cluster. 
 \emph{Biometrics}, \textbf{67}, 629--635.

 \bibitem[Liang and Zeger(1986)]{liang1986}
 Liang, K.-Y., and Zeger, S. L. (1986).
 Longitudinal data analysis using generalized linear models. 
 \emph{Biometrika}, \textbf{73}, 13--22.
 
 \bibitem[Little and Rubin(2002)]{little2002}
 Little, R. J. A., and Rubin, D. A. (2002).
 \emph{Statistical Analysis with Missing Data, 2nd edition}.
 New Jersey: John Wiley \& Sons.

 \bibitem[Liu and Hedeker(2006)]{liu2006}
 Liu, L. C., and Hedeker, D. (2006).
 A mixed-effects regression model for longitudinal 
 multivariate ordinal data. 
 \emph{Biometrics}, \textbf{62}, 261--268.

 \bibitem[Molenberghs et~al.(1997)]{molenberghs1997}
 Molenberghs, G., Kenward, M. G., and Lesaffre E. (1997).
 The analysis of longitudinal ordinal data with 
 nonrandom drop-out. 
 \emph{Biometrika}, \textbf{84(1)}, 33--44.

 \bibitem[Neal(2011)]{neal2011}
 Neal, R. (2011). MCMC using Hamiltonian dynamics. In Brooks, St., Gelman, A., Jones, G. L., and Meng, X.L. eds. 
 Handbook of Markov Chain Monte Carlo, pages 113--162.
 Boca Raton: Chapman \& Hall/CRC Press.

 \bibitem[Nooraee et~al.(2016)]{nooraee2016}
 Nooraee, N., Abegaz, F., Ormel, J., Wit, E., and 
 van den Heuvel E. R. (2016).
 An approximate marginal logistic distribution for the 
 analysis of longitudinal ordinal data. 
 \emph{Biometrics}, \textbf{72}, 253--261.

 \bibitem[Oguz-Alper and Berger(2015)]{oguzalper2015}
 Oguz-Alper, M., and Berger, Y. G. (2015). 
 Variance estimation of change in poverty rates: an application to the Turkish EU-SILC survey. 
 \emph{Journal of Official Statistics}, \textbf{31(2)}, 155--175.

 \bibitem[Parzen et~al.(2011)]{parzen2011}
 Parzen, M., Ghosh, S., Lipsitz, S., Sinha, D., Fitzmaurice, G. M., Mallick, B. K., and Ibrahim J. G. (2011). 
 A generalized linear mixed model for longitudinal binary data with a marginal logit link function. 
 \emph{The Annals of Applied Statistics}, \textbf{5(1)}, 449--467.

 \bibitem[Perin et~al(2014)]{perin2014}
 Perin, J., Preisser, J. S., Philips, C., and Qaqish, B. (2014).
 Regression Analysis of Correlated Ordinal Data Using
Orthogonalized Residuals. 
 \emph{Biometrics}, \textbf{70(4)}, 902--909.

 \bibitem[Pirani and Salvini(2015)]{pirani2015}
 Pirani, E., and Salvini, S. (2015).
 Is temporary employment damaging to health? a longitudinal study on Italian workers. 
 \emph{Social Science \& Medicine}, \textbf{124}, 121--131.

 \bibitem[Polson and Scott(2012)]{polson2012}
 Polson, N. G., and Scott, J. G. (2012)
 On the Half-Cauchy Prior for a Global Scale parameter. 
 \emph{Bayesian Analysis}, \textbf{7(4)}, 887--902.

  \bibitem[Pulkstenis et~al.(2001)]{pulkstenis2001}
 Pulkstenis, E., Ten Have, T. R., and Landis J. R. (2001). 
 A mixed effects model for the analysis of ordinal longitudinal 
pain data subject to informative drop-out. 
 \emph{Statistics in Medicine}, \textbf{20}, 601--622.

 \bibitem[R Core Team(2019)]{r2019}
 R Core Team (2019).
 \emph{R: A language and environment for statistical computing}. 
 R Foundation for Statistical Computing, Vienna, Austria. 
 URL https://www.R-project.org/.

 \bibitem[Raman and Hedeker(2005)]{raman2005}
 Raman, R., and Hedeker, D. (2005). 
 A mixed-effects regression model for three-level ordinal response data. 
 \emph{Statistics in Medicine}, \textbf{24}, 3331--3345.

 \bibitem[Rana et~al(2018)]{rana2018}
 Rana, S., Roy, S., Das, K., for the Alzheimer’s Disease
Neuroimaging Initiative (2018). 
 Analysis of ordinal longitudinal data under nonignorable 
missingness and misreporting: an application to Alzheimer’s
disease study. 
 \emph{Journal of Multivariate Analysis}, \textbf{166}, 62--77. 

 \bibitem[Reeves et~al.(2014)]{reeves2014}
 Reeves, A., Karanikolos, M., Mackenbach, J., McKee, M., and Stuckler, D. (2014). 
 Do employment protection policies reduce the relative disadvantage 
 in the labour market experienced by unhealthy people? A natural 
 experiment created by the Great Recession in Europe. 
 \emph{Social Science \& Medicine}, \textbf{121}, 98--108.


 \bibitem[Stan Development Team(2018)]{rstan2018}
 Stan Development Team (2018).
 \emph{{RStan}: the {R} interface to {Stan}}. 
 R package version 2.17.3. URL http://mc-stan.org/.
 
 \bibitem[Tom et~al.(2015)]{tom2016}
 Tom, B. D. M., Su, L., and Farewell, V. T. (2016). 
 A corrected formulation for marginal inference derived from two-part mixed models for longitudinal semi-continuous data. 
 \emph{Statistical Methods in Medical Research}, \textbf{25(5)}, 2014--2020.

 \bibitem[T{\o}ge and Blekesaune(2015)T{\o}ge and Blekesaune]{toge2015}
 T{\o}ge, A. G., and Blekesaune, M. (2015). 
 Unemployment transitions and self-rated health in Europe: A longitudinal analysis of EU-SILC from 2008 to 2011.
 \emph{Social Science \& Medicine}, \textbf{143}, 171--178.

 \bibitem[T{\o}ge(2016)]{toge2016}
 T{\o}ge, A. G. (2016). 
 Health effects of unemployment in Europe (2008--2011): A longitudinal analysis of income and financial strain as mediating factors. 
 \emph{International Journal for Equity in Health}, \textbf{15(75)}, 1--12.

 \bibitem[Tu et~al.(2011)]{tu2011}
 Tu, W., Ghosh, P., and Katz, B. P. (2011). 
 A stochastic model for assessing Chlamydia trachomatis 
 transmission risk by using longitudinal observational data. 
 \emph{Journal of the Royal Statistical Society - Series A}, \textbf{174(4)}, 975--989.

 \bibitem[Vaalavuo(2016)]{vaalavuo2016}
 Vaalavuo, M. (2016). 
 Deterioration in health: What is the role of unemployment and poverty? 
 \emph{Scandinavian Journal of Public Health}, \textbf{44(4)}, 347--353.

 \bibitem[van der Wel et~al.(2011)]{vanderwel2011}
 van der Wel, K. A., Dahl, E., and Thielen K. (2011). 
 Social inequalities in `sickness': European welfare states and non-employment among the chronically ill. 
 \emph{Social Science \& Medicine}, \textbf{73}, 1608--1617.

  \bibitem[van Steen et~al.(2001)]{vansteen2001}
 van Steen, K., Molenberghs, G., Verbeke, G., and Thijs, H. (2001). 
A local influence approach to sensitivity analysis of
incomplete longitudinal ordinal data. 
 \emph{Statistical Modelling}, \textbf{1}, 125--142.

 \bibitem[Varin and Czado(2010)]{varin2010}
 Varin, C. and Czado, C. (2010). 
A mixed autoregressive probit model for ordinal
longitudinal data. 
 \emph{Biostatistics}, \textbf{11(1)}, 127--138.

 \bibitem[Wang and Louis(2003)]{wang2003}
 Wang, Z., and Louis T. A. (2003). 
 Matching conditional and marginal shapes in binary random intercept models using a bridge distribution function. 
 \emph{Biometrika}, \textbf{90(4)}, 765--775.

 \bibitem[Wang and Louis(2004)]{wang2004}
 Wang, Z., and Louis, T. A. (2004). 
 Marginalized binary mixed-effects models with covariate-dependent random effects and likelihood inference. 
 \emph{Biometrics}, \textbf{60}, 884--891.

 \bibitem[Watanebe(2010)]{watanebe2010}
 Watanebe, S. (2010). 
 Asymptotic equivalence of Bayes cross validation and widely applicable information criterion in singular learning theory. 
 \emph{Journal of Machine Learning Research}, \textbf{11}, 3571--3594.

 \bibitem[Yardim and Uner(2018)]{yardim2018}
 Yardim, M. S., and Uner S. (2018). 
 Equity in access to care in the era of health system reforms in Turkey.
 \emph{Health Policy}, https://doi.org/10.1016/j.healthpol.2018.03.016.

  \bibitem[Zaloumis et~al.(2015)]{zaloumis2015}
 Zaloumis, S. G., Scurrah, K. J., Harrap, S. B., 
 Ellis, J. A., and Gurrin L. C. (2015). 
Non-proportional odds multivariate logistic regression of ordinal
family data. 
 \emph{Biometrical Journal}, \textbf{57}, 286--303.

\end{thebibliography}
\end{document}